\documentclass[usenatbib,fleqn]{mnras}
\usepackage{graphicx,hyperref,amsmath,amssymb,mathrsfs,xspace}
\usepackage{soul}
\usepackage[dvipsnames]{xcolor}

\newcommand{\ENZO}{\texttt{Enzo}\xspace}
\newcommand{\Grackle}{\texttt{Grackle}\xspace}

\title[Population III Star Formation]{The Role of Radiation and Halo Mergers in Pop III Star Formation}

\author[Correa Magnus et al.]{
Lilia Correa Magnus,$^{1}$
Britton D. Smith,$^{1}$
Sadegh Khochfar,$^{1}$
Brian W. O'Shea, $^{2,3,4,5}$
\newauthor
John H. Wise, $^{6}$
Michael L. Norman, $^{7,8}$ 
and Matthew J. Turk $^{9,10}$ 
\\
$^{1}$Institute for Astronomy, University of Edinburgh, Royal Observatory, Edinburgh EH9 3HJ, UK\\
$^{2}$Department of Physics and Astronomy, Michigan State University, East Lansing, MI 48824, USA\\
$^{3}$Department of Computational Mathematics, Science, and Engineering, Michigan State University, East Lansing, MI 48824, USA\\
$^{4}$National Superconducting Cyclotron Laboratory, Michigan State University, East Lansing, MI 48824, USA\\
$^{5}$JINA-CEE: Joint Institute for Nuclear Astrophysics-Center for the Evolution of the Elements, USA\\
$^{6}$Center for Relativistic Astrophysics, School of Physics, Georgia Institute of Technology, Atlanta, GA, 30332, USA\\
$^{7}$San Diego Supercomputer Center, University of California, San Diego, 10100 Hopkins Drive, La Jolla, CA 92093, USA\\
$^{8}$Center for Astrophysics and Space Sciences, University of California, San Diego, 9500 Gilman Dr, La Jolla, CA 92093, USA\\
$^{9}$School of Information Sciences, University of Illinois, Urbana-Champaign, IL, 61820, USA\\
$^{10}$Department of Astronomy, University of Illinois, Urbana-Champaign, IL, 61820, USA
}

\pubyear{2023}

\begin{document}
\label{firstpage}
\pagerange{\pageref{firstpage}--\pageref{lastpage}}
\maketitle

\begin{abstract}

We present a study of the co-evolution of a population of primordial star-forming minihalos at Cosmic Dawn. In this study, we highlight the influence of individual Population III stars on the ability of nearby minihalos to form sufficient molecular hydrogen to undergo star formation. In the absence of radiation, we find the minimum halo mass required to bring about collapse to be $\sim$10$^5\ \mathrm{M_{\odot}}$, this increases to $\sim$10$^6\ \mathrm{M_{\odot}}$ after two stars have formed. We find an inverse relationship between halo mass and the time required for it to recover its molecular gas after being disrupted by radiation from a nearby star. We also take advantage of the extremely high resolution to investigate the effects of major and minor mergers on the gas content of star-forming minihalos. Contrary to previous claims of fallback of supernova ejecta, we find minihalos evacuated after hosting Pop III stars primarily recover gas through mergers with undisturbed halos. We identify an intriguing type of major merger between recently evacuated halos and gas-rich ones, finding that these ``mixed" mergers accelerate star formation instead of suppressing it like their low redshift counterparts. We attribute this to the gas-poor nature of one of the merging halos resulting in no significant rise in temperature or turbulence and instead inducing a rapid increase in central density and hydrostatic pressure. This constitutes a novel formation pathway for Pop III stars and establishes major mergers as potentially the primary source of gas, thus redefining the role of major mergers at this epoch. 
\end{abstract}




 
\section{Introduction} \label{sec:intro}


The first stars to form in the universe, classified as Population III (Pop III), triggered an important transition in the early evolution of structure formation. The environment at this epoch, being mostly composed of hydrogen and helium -- greatly contrasts the diverse and increasingly complex universe observed at present (see reviews by \citealt{Ciardi}; \citealt{Glover_2012} with \citealt{Bromm_2013}). The birth of these first luminous objects initiated reionisation and their supernovae enriched the intergalactic medium (IGM) and enabled the formation of subsequent generations of stars. Understanding the conditions that lead to their creation is therefore crucial to fully comprehend the Cosmic Dawn era and the seeding of the first galaxies.

The discussion of this epoch also requires a brief delve into the current concordance model to establish some cosmological context. Under the $\Lambda$CDM framework, dark matter is assumed to be cold and weakly interacting. This introduces a hierarchical growth system whereby dark matter halos grow from small to large via a series of mergers (\citealt{Couchman1986}; \citealt{Khochfar_2008}). The halos that hosted Pop III stars are therefore assumed to have been small -- usually referred to as minihalos -- and had virial temperatures of around $10^2-10^3$ K. Studies such as \citet{Couchman1986} and \citet{Tegmark1997} established these initial virial temperatures and highlighted the unique scenario minihalos presented in the formation of Pop III stars. The environment provided by these young halos directly  determined the formation channels available to the first stars. The metal-free gas present alongside virial temperatures below the atomic hydrogen cooling limit meant a cooling mechanism was required for gas to effectively collapse. At high redshifts, the dominant cooling channel in primordial gas is via molecular hydrogen transitions (\citealt{Peebles}; \citealt{Omukai_2001}). \cite{Tegmark1997} also presented analytic methods to model this early collapse period alongside analysis of the chemical processes involved in the production of $\mathrm{H_2}$. Because minihalos do not readily contain enough $\mathrm{H_2}$ to efficiently cool, they first require a period of $\mathrm{H_2}$ build-up. At high redshifts, $\mathrm{H_2}$ is created primarily via the $\mathrm{H^-}$ channel ($\mathrm{H} + \mathrm{H^-} \rightarrow \mathrm{H_2} + \mathrm{e^-}$) and the $\mathrm{H_{2}^+}$ channel ($\mathrm{H_{2}^+} + \mathrm{H} \rightarrow \mathrm{H_2} + \mathrm{H^+}$). Studies of this early period (\citealt{Abel2002}; \citealt{Yoshida2003}; \citealt{Reed_2005}; \citealt{Wise_2007}; \citealt{Oshea2007}; \citealt{Yajima_2017}) generally agree that the first stars formed at $z\sim20-30$, once the first halos reached a high enough $\mathrm{H_2}$ fraction. The fragile nature of $\mathrm{H_2}$, however, means this process is not as straightforward at slightly later epochs. Molecular hydrogen is easily photo-dissociated by UV radiation in the Lyman-Werner (LW) bands (11.2-13.6 eV), which is emitted by Pop III stars and can reach the cores of nearby halos if these are not yet large enough for self-shielding to take place (\citealt{Oshea2008}; \citealt{Agarwal_2015}). The presence of nearby star formation can therefore set-back collapse by lowering $\mathrm{H_2}$ levels below what is needed to cool efficiently (\citealt{jonhson_2013}; \citealt{Agarwal_2019}).

Over the past 20 years cosmological simulations have been used to study this high redshift period and works have focused heavily on the effect of LW radiation has on the halo mass scale of star forming minihalos. Early research work such as \cite{Tegmark1997}, \cite{Machacek_2001} and \cite{Yoshida2003} attempted to  establish a function for the critical halo mass $\mathrm{M_{crit}}$ above which minihalos are able to form stars. They also established LW radiation as an effective suppressant of star formation, agreeing on the heavy impact this type of feedback had, pushing $\mathrm{M_{crit}}$ up from $\sim10^5$ $\mathrm{M_{\odot}}$ under no radiation to $10^6$ $\mathrm{M_{\odot}}$ or even $10^7$ $\mathrm{M_{\odot}}$ for irradiated halos. More recent studies such as \cite{Park_2021}, \cite{Schauer2021}, \cite{Kulkarni_2021} and \cite{Incatasciato_2023} have continued to investigate this topic, with additional considerations such as an X-ray background or baryon-dark matter streaming velocities alongside LW radiation. Although more nuanced -- due to the additional factors considered -- these works also found delayed collapse scenarios in environments where halos were exposed to UV photons. The evolution of the $\mathrm{M_{crit}}$ function is still a debated topic, with many finding a redshift dependence: \cite{Tegmark1997}; \cite{Trenti_2009}; \cite{Kulkarni_2021}; and others contesting this claim: \cite{Machacek_2001}; \cite{Schauer2019a}; \cite{Schauer2021} and \cite{Incatasciato_2023}. Despite this tension, there is a strong consensus that LW radiation effectively suppresses and delays star formation at early epochs by lowering the $\mathrm{H_2}$ content inside minihalos. However, these works have focused on the large-scale environmental impact of UV radiation via a steadily building LW background. Little attention has been paid to the smaller scale impact of discrete populations of nearby Pop III stars.

Building on the chemical model presented in \cite{Tegmark1997}, some works have also attempted to account for other coolants such as $\mathrm{HD}$ (\citealt{Yoshida_2006}; \citealt{Ripamonti_2007}; \citealt{Greif_2008}). The advantage of cooling via HD is its asymmetry, which provides a permanent dipole moment and allows gas to cool down to CMB temperatures. This cooling channel contrasts the $\mathrm{H_2}$ rotational transitions, which become inefficient below $\sim$200 K, and above number densities of $\sim$10$^{3-4}$ cm$^{-3}$, when populations are dictated by local thermodynamic equilibrium (LTE). As discussed in studies that looked at molecule formation in the early universe such as \cite{Galli98} and \cite{Stancil98}, production of HD under Cosmic Dawn conditions is only possible via the enhancement of the $\mathrm{H_2}$ formation reaction itself. Given the reaction: $\mathrm{H_2} + \mathrm{D^+} \rightarrow \mathrm{HD} + \mathrm{H^+}$, gas clouds that contained a higher HD fraction at this epoch are also expected to be highly ionized. Based on this idea, \cite{Shchekinov_2005} produced a simplified one-dimensional model to find whether mergers could boost the $\mathrm{H^-}$ and $\mathrm{e^-}$ content, helping catalyse the formation of 
$\mathrm{H_2}$ and $\mathrm{HD}$. Later studies such as \cite{Prieto_2012} and \cite{Bovino2014}, would expand on this by producing more realistic merger scenarios in hydro-cosmological simulations. They both found turbulence and shock-waves from such dynamical events could increase HD levels and provide an alternative pathway to star formation.

These theories could initially be seen as contradictory to studies that assert dynamical events such as major mergers delay star formation (\citealt{Yoshida2003}; \citealt{Fernandez_2014}; \citealt{Wise_2019}). Starting from the Rees-Ostriker relation \citep{rees}: $t_\mathrm{dyn} \gg t_\mathrm{cool}$; which requires cooling timescales to be much smaller than dynamical times for collapse to occur, any mechanism that can invert this relation results in the opposite effect.  Fast accretion, in the form of major mergers provides such a scenario. Increased turbulence and shock-heating from these events is thought to be able to greatly reduce dynamical timescales and therefore suppress star formation. \cite{Wise_2019} find important delays in their star forming halos, as turbulence maintains gas pressure-supported against collapse. Moreover, \cite{Latif_2022} find that the cosmic web collapse provides enough turbulence to prohibit star formation, even at moderate growth rates of halos. In their research of HD cooling,  \cite{Prieto_2012} and \cite{Bovino2014} do see a large increase in gas turbulence and heating, which is required to increase the abundance of $\mathrm{H_2}$ and boost the HD content. Although these works don't specifically mention delays in star formation, none of their collapse scenarios occur whilst a major merger is ongoing. Further studies on the subject such as \cite{Hirano_2014} additionally suggest a slow collapse mode where gas clouds are rotationally supported can provide the right conditions for HD cooling to become effective. Both concepts are not mutually exclusive and could occur in the same merger events, though the possibility of HD enhancement adds complexity to the effects of major mergers on Pop III star formation.

While there is general consensus that Pop III star formation is heavily influenced by environmental factors (i.e., radiation and structure formation), few works have investigated the scenario of minihalos co-evolving amidst the formation of discrete stellar populations. In this paper, we study the conditions that lead to the formation of metal-free, star-forming gas with a cosmological simulation that includes the radiative feedback of individual Pop III stars. The simulation resolves halos that are many orders of magnitude smaller than the canonical minimum halo mass for metal-free star formation, thus allowing us to robustly probe the influence of minor mergers. The paper is structured as follows: the details of the simulation code and setup are introduced in Section \ref{sec:code}, our results are presented in Section \ref{sec:results}, followed by a discussion alongside the implication of our findings in Section \ref{sec:discussion}. Our final summary and conclusions are then discussed in Section \ref{sec:conclusion}.

\section{Methods}
\label{sec:code}
The simulation analyzed in this work was first presented in \citet{Smith2022}. It is the second in the \texttt{Pop2Prime} series of simulations, first presented in \citet{Britton2015}. We refer readers to that work for a detailed discussion of the methods employed. The simulation was run using the open-source, adaptive mesh-refinement + N-body code \ENZO \citep{enzo,Brummel-Smith2019}. \ENZO has been used extensively to study Pop III star formation at high redshift \citep[e.g.,][]{Abel2002, Oshea2007, Turk2009, Bovino2014, Skinner_2020}. A more thorough explanation of this framework is presented in \cite{enzo}, although our overview should be enough to cover the functionality most relevant to our work. The analysis of the simulation snapshots and merger trees was performed using the \texttt{yt} \citep{yt} and \texttt{ytree} \citep{ytree} packages. Halo finding and merger tree generation were performed with the \texttt{Rockstar} halo finder \citep{2013ApJ...762..109B} and \texttt{consistent-trees} \citep{2013ApJ...763...18B}.

\subsection{\ENZO}
\ENZO is a hydrodynamics and N-body cosmological simulation code that uses adaptive mesh-refinement (AMR) to model structure formation under the $\Lambda$CDM framework. It uses AMR to introduce varying levels of refinement within the simulation grid depending on the amount matter content and detail required in the volume \citep{Berger98}. As a result, the coarseness of the grids is either refined or de-refined such that the grid is increasingly subdivided to obtain higher detail in the denser regions. Within this basis, hydrodynamics and adaptive particle-mesh gravity solvers are implemented, in this case the Piecewise Parabolic Method from \cite{woodward98} is used alongside an N-body adaptive particle-mesh gravity solver \citep{efstathiou85}. Finally, when a radiation source is present in the volume, the \textsc{Moray} adaptive radiation transport method from \cite{WiseandAbel2011} is applied to simulate the radiating field.

We make use of the precursor to the \Grackle library \citep{Grackle}, built-in to the \ENZO code, to follow non-equilibrium primordial chemistry and cooling for the $\mathrm{e^-}$, $\mathrm{H}$, $\mathrm{D}$, and $\mathrm{He}$ species as well as their ionic and molecular variants: $\mathrm{H^+}$, $\mathrm{H^-}$, $\mathrm{H_2}$, $\mathrm{H_2^+}$, $\mathrm{D}$, $\mathrm{D^+}$, $\mathrm{He^+}$, $\mathrm{He^{++}}$, $\mathrm{HD}$. We also include the cooling from heavy elements with tables of rates pre-computed with the \texttt{Cloudy} \citep{2013RMxAA..49..137F}, which assume a solar abundance pattern and collisional ionization equilibrium \citep{2008MNRAS.385.1443S}.

We simulate the formation of individual Pop III stars using the method from \cite{wise2012}, where dense gas is followed until it reaches the criteria detailed below. The criteria for Pop III star formation is exactly that detailed in \cite{Britton2015} except with a higher metallicity threshold of 10$^{-4}$ Z$_{\odot}$, instead of 10$^{-6}$ Z$_{\odot}$. The specific star particle creation criteria are as follows:
\begin{enumerate}
\item The proper  baryon number density,  $n_\mathrm{b} > 10^7 \mathrm{cm}^{-3}$.
\item The gas flow is convergent ($\nabla \cdot v_\mathrm{gas} < 0$).
\item The molecular hydrogen mass fraction; $f_\mathrm{H_2}$ is greater \indent \indent \indent  than $5 \times 10^{-4}$, where $f_\mathrm{H_2} \equiv (\rho_\mathrm{H_2}+\rho_\mathrm{H_2^+})/\rho_\mathrm{b}$
\item The metallicity is below $10^{-4}$ $\mathrm{Z_{\odot}}$ 
\end{enumerate}

\noindent All Pop III star particles produced in the simulation are identical, with masses of $\mathrm{M} = 40$ $\mathrm{M_{\odot}}$ and lifetimes of $3.86$ $\mathrm{Myr}$, and all ending as core-collapse supernovae with energies of $10^{51}$ ergs. These properties were taken from the stellar evolution models presented in \cite{Schaerer2002}. Each star is treated as a point source that radiates from the moment it is placed in its halo. The ionizing radiation from the star particles is modelled using the \texttt{Moray} adaptive ray-tracing method and following 3 energy groups that ionize: H with $\mathrm{E} = 28$ eV and $\mathrm{L_{\gamma}} = 2.47 \times 10^{49}\ s^{-1}$; He with $E = 30$ eV and $\mathrm{L_{\gamma}} = 1.32 \times 10^{49}\ s^{-1}$; $\mathrm{He^+}$ with $E = 58$ eV and $\mathrm{L_{\gamma}} = 8.80 \times 10^{46}\ s^{-1}$. $\mathrm{H_2}$ photo-dissociating radiation is also emitted with $\mathrm{L_{\gamma}} = 2.90 \times 10^{49}\ s^{-1}$ and modelled to decline as $\mathrm{r^{-2}}$ alongside the $\mathrm{H_2}$ ``Sobolev-like'' self-shielding model from \citet{Wolcott_Green}. The unique approach in this simulation combined the absence of a uniform background radiation with a high resolution capacity, allowed us to treat the formation of each star as an individual radiative event. 

\subsection{Simulation Environment}
A 500 kpc/h simulation box was initialized using the \textsc{music} initial conditions \citep{MUSIC} with the \textsc{WMAP-7} \citep{wmap} cosmological parameters $\Omega_\mathrm{m} = 0.266$, $\Omega_\mathrm{\Lambda} = 0.732$, $\Omega_\mathrm{b} = 0.0449$, $\mathrm{H_0} = 71.0$ km/s/Mpc, $\sigma_8 = 0.801$, and $n_\mathrm{s} = 0.963$. The simulation zooms in on a halo reaching 10$^{7}$ M$_{\odot}$ at $z \sim 10$ with 512$^{3}$ cells/particles and two initial levels of refinement, for an effective resolution of $2048^3$. This corresponds to an initial spatial and dark matter resolution of 0.244 $\mathrm{kpc/h}$ and $\sim 1$ $\mathrm{M_{\odot}}$, respectively. The simulation was designed to model the conditions leading to the formation of the first Pop II stars assuming the original critical metallicity based on gas-phase metal cooling \citep{2003Natur.425..812B}. When dense gas above this metallicity forms, the simulation comes to an end. This occurs at $z\sim11.8$. The simulation is the exact same realization presented in \citet{Britton2015}, only with the higher threshold metallicity, allowing it to run for an additional $\sim$250 Myr. We find a total of 12 Pop III stars within 9 distinct halos, all of which are studied for this paper. 

\section{Results}
\label{sec:results}

\begin{table*}
	\centering
	\begin{tabular}{| c | c | c | c | c | c |}
		\hline
		  Halo label & Formation Time (Myr)& Halo Mass ($\mathrm{M_{\odot}}$) & Double star (Y/N) & Metallicity ($\mathrm{Z_{\odot}}$) & Major mergers (\#) \\ & & & & & (mixed merger before star? Y/N) \\
		\hline
		Halo 1 & 144 & $1.2\times10^5$ & No & 0 & 0 (N) \\
        Halo 2 & 211 & $2.0\times10^5$ & No & 0 &0 (N) \\
        Halo 3 & 240 & $9.8\times10^5$ & No & $2.6\times10^{-5}$ &3 (Y) \\
        Halo 4 & 249 & $5.2\times10^5$ & No & 0 &2 (N)\\
        Halo 5/6 & 279 & $1.1\times10^6$ & Yes & 0 & 1 (N) \\
        Halo 7/8 & 289 & $1.8\times10^6$ & Yes & 0 & 1 (N)\\
        Halo 9 & 312 & $8.6\times10^5$ & No & 0 & 0 (N) \\
        Halo 10/11 & 344 & $6.8\times10^5$ & Yes & 0 & 1 (N)\\
        Halo $12^*$ & 379 & $7.3\times10^6$ & No & $5.2\times10^{-5}$& 4 (Y) \\
		\hline
	\end{tabular}
	\caption{Table detailing the star formation times, halo mass, halos that hosted two stars, metallicity within the densest gas cell before star formation, number of major mergers that occurred before a star was created and whether a mixed merger occurred before star formation for each of the halos.}
	\label{tab:halo_info}
\end{table*}

Before delving into the results, we define the following the halo labelling system. Each minihalo will henceforth be referred to by the star it formed such that Halo 1 is the halo that formed the first star, Halo 2 the second and so on. Three of the halos produced two stars a few thousand years apart; in this case the $5^{\mathrm{th}}$ and $6^{\mathrm{th}}$ stars formed inside the same halo, as did the $7^{\mathrm{th}}$ and $8^{\mathrm{th}}$ stars and $10^{\mathrm{th}}$ and $11^{\mathrm{th}}$ stars. We treat them as the same radiative event due to their proximity in space and time and no distinction between the stellar objects is made in our analysis, we therefore label these minihalos as Halos 5/6, 7/8 and 10/11. We also give a special label to the halo that forms the final star: Halo $12^*$, since it is the same halo that previously formed the $3^{rd}$ star. Table \ref{tab:halo_info} provides an overview of the star formation events and their associated halos. We note the occurrence of three events in which two star particles formed in the same halo within a few thousand years of each other. We did not identify any obvious distinguishing characteristics of these halos, but will examine this further in a future work.

\begin{figure}
    \centering
    \includegraphics[width=\linewidth]{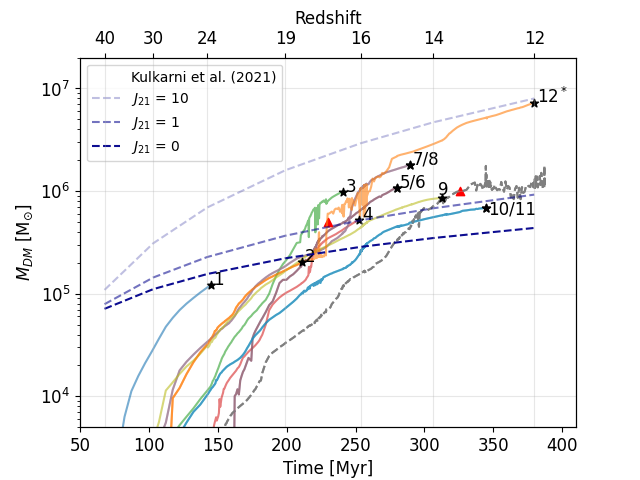}
    \caption[width=\linewidth]{Mass growth of halos from the beginning of the simulation up to the moment they formed a star, labelled with the halo they formed in. The grey dashed line shows the evolution of the halo that forms the final star. Because it merges into Halo 2 (orange line) the halo finder assumes it is formed there instead, so the final star is labeled as $12^*$. The initial time of the two mixed major mergers that occur in the simulation are marked with a red triangle. Overplotted in blue are fits of $\mathrm{M_{crit}}$ for different levels of background radiation ($\mathrm{J_{21}}$) assuming no streaming velocities, taken from \citet{Kulkarni_2021}.}
    \label{fig:growth}
\end{figure}

\subsection{The Impact of Radiation}

Past works have discussed the clear connection between radiative feedback and star formation, with photo-dissociation reducing or even halting the cooling of gas inside halos. We therefore investigated $\mathrm{H_2}$ content alongside halo growth to better understand when radiation effectively delays star birth during halo evolution.

In Figure \ref{fig:growth} we present the mass growth of the star forming minihalos starting from very low masses at z = $30-40$ up to the point of star formation. The first halo to undergo star formation is also the first to reach a mass of $10^5$ $\mathrm{M_{\odot}}$ at $z\sim24$. Halo 1 experiences no radiation feedback during this period and undergoes no major or minor mergers; it is therefore an example of star formation under no external influence. Halo 2 is the only other halo that forms a star while still having a relatively small mass: $2\times10^5 \mathrm{M_{\odot}}$, although it undergoes a small merger event that introduces a significant amount of gas before its star forms ($\sim$1/4 of its total gas content at $\sim$150 Myr). After two sets of radiation bursts, none of the other halos form a star near this mass threshold; instead, they undergo star birth once they approach a mass of $\sim$10$^6$ $\mathrm{M_{\odot}}$ with the final star forming inside one of the largest halos in the system: $7\times10^6$ $\mathrm{M_{\odot}}$. 

Figure \ref{fig:growth} leads to a great amount of insight into the effects of radiation, as we find that only two events are needed to increase the mass of a star forming halo by an order of magnitude. Subsequent stars therefore take longer to form and Halos $3-11$ experience close star formation events as they all approach this higher mass threshold at similar times. Another unique scenario is also reflected in this figure: the last star that forms inside Halo 3, which had previously merged with Halo 2, technically forms inside a subhalo structure created after a major merger. This merging halo has a shallower growth curve, as seen from the dashed line, since it is the last halo to reach a mass of $10^5$ $\mathrm{M_{\odot}}$ we naturally expect it to form a star at a much later epoch than the rest of the halos.

\begin{figure}
    \centering
    \includegraphics[width=\linewidth]{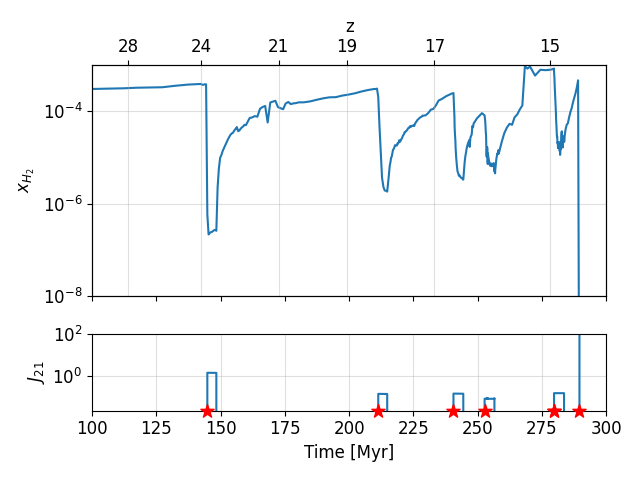}
    \caption[width=\linewidth]{Evolution of the $\mathrm{H_2}$ fraction alongside the intensity of the LW radiation in units of $10^{-21}$ erg $\mathrm{s^{-1}\ cm^{-2}\ sr^{-1}\ Hz{^{-1}}}$ at the densest point in Halo 7/8's gas core. The red stars in the x-axis mark the time of each star's birth.}
    \label{fig:h2_frac}
\end{figure}

It should be noted, that despite Halo 3 also reaching $\mathrm{M_{crit}}$ at a similar time to Halo 2 (see Figure \ref{fig:growth}) it would take an extra $20$ Myr for it to form a star. We attribute this to its accelerated growth slowing the efficiency of its cooling enough for Halo 2 to form a star first. Radiation feedback and the blastwave from the second star further exacerbated this delay, but we will delve into Halo 3's unique evolution with more detail in Section \ref{sec:mergers}. 

The standard effect of LW radiation on our halos is shown in Figure \ref{fig:h2_frac}. It presents the $\mathrm{H_2}$ fraction measured at the densest gas cell in Halo $7/8$ up until the formation of its stars. Figure \ref{fig:h2_frac} also highlights how the simulation has no background radiation and instead models individual ``bursts" produced by the stars. Halos are generally exposed to $\mathrm{J_{21}}$ values of $\sim0.1$ to 1 while other stars are shining. By analyzing this Figure, we find that $\mathrm{H_2}$ fractions are more heavily reduced when the mass of the halo is still below or near $10^5$ $\mathrm{M_{\odot}}$. We can see however, that the effectiveness of photo-dissociation diminishes with time, with faster recoveries following each stellar event. Halo $7/8$ (light purple line in Figure \ref{fig:growth}) takes about $100$ Myr to grow from $10^5$ $\mathrm{M_{\odot}}$ to $10^6$ $\mathrm{M_{\odot}}$, throughout this time, the amount of $\mathrm{H_2}$ dissociated by each star that forms in the system diminishes as does the time taken to recover back to star forming levels. This indicates a clear relation between halo growth and the total delay caused by nearby star formation. Although only Halo $7/8$ is included for clarity, similar $\mathrm{H_2}$ fraction curves are observed for all halos that form stars.

\begin{figure}
    \centering
    \includegraphics[width=\linewidth]{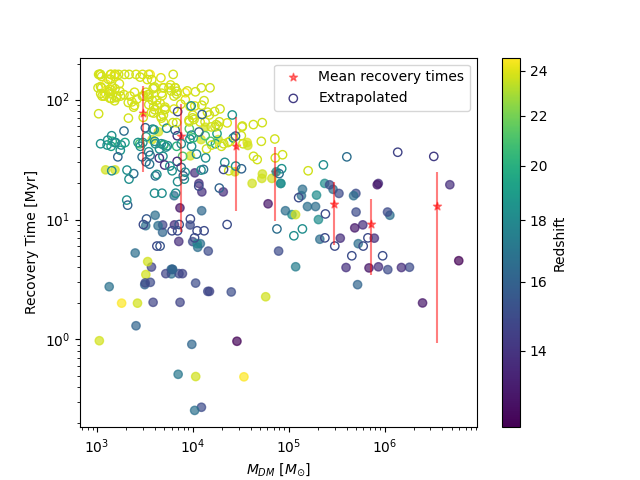}
    \caption[width=\linewidth]{Halo $\mathrm{H_2}$ recovery times against their dark matter mass at times just after star formation for halos that met the criteria. Extrapolated points are plot as open circles. Mean recovery times for unevenly spaced mass bins: $10^3 - 5\times10^3$; $5\times10^3 - 10^4$; $10^4 - 5\times10^4$; $5\times10^4 - 10^5$; $10^5 - 5\times10^5$; $5\times10^5 - 10^6$; $10^6 - 10^7$; and their standard deviations are plot in red.}
    \label{fig:recovery}
\end{figure}

To better quantify the relation between photo-dissociation and halo growth we looked more closely at recovery times. In Figure \ref{fig:recovery} we present the calculated and estimated recovery times for halos in the simulation. The halos included in this plot had to fullfill the following criteria:

\begin{enumerate}
    \item Within the halo's progenitor line, the halo at some point in its evolution must exceed a dark matter mass of $10^4$$\mathrm{M_{\odot}}$.
    
    \item At the time of a radiative event, the halo must have a molecular hydrogen fraction $\mathrm{f_{H_2}} \ge 10^{-4}$ within its densest grid cell and a dark matter to baryon mass ratio above 1\% of the cosmic mean ($\Omega_\mathrm{baryon}\ /\ \Omega_\mathrm{DM}$).

    \item The halo must not be a subhalo.
    
\end{enumerate}

With this criteria we attempted to remove all the subhalos, gas poor halos and halos that didn't have a high enough  $f_\mathrm{H_2}$ to be recovered in the first place. To measure the recovery time itself we found the highest $\mathrm{H_2}$ fraction reached by each halo before a star formed in the system. We measured the $\mathrm{H_2}$ fraction in the densest cell of each halo since this is where we would expect star formation to occur. We defined this as $\mathrm{f_{H_2}^{\ 0}}$ and found the amount of time (after the star's death) it took the halos to recover half of their original peak molecular hydrogen content: $\frac{1}{2}\mathrm{f_{H_2}^{\ 0}}$. The halo mass was taken at the time of $\mathrm{f_{H_2}^{\ 0}}$, as this established the initial $\mathrm{H_2}$ production rate. Many of the halos never managed to reach $\frac{1}{2}\mathrm{f_{H_2}^{\ 0}}$ before the next star formed, in these cases we approximated the recovery curve to a logarithmic function since the form of $f_\mathrm{H_2} (t) $ is well fit by this as can be seen from the recovery curves in Figure \ref{fig:h2_frac}. We therefore made use of the function in equation \ref{eq:log} to find the extrapolated times.
\begin{equation}
\label{eq:log}
   f_\mathrm{H_2}(t) = A+Blog_{10}(Ct) + Dt
\end{equation}

Where A, B, C and D were free parameters to be fit. The parameters found for each halo varied due to the redshift, mass and accumulated radiation events. Using equation \ref{eq:log}, we estimate the time it would have taken these halos to reach $\frac{1}{2}\mathrm{f_{H_2}^{\ 0}}$ had the subsequent radiative event not occurred. The detriment of using such a simple model is evident in Figure \ref{fig:recovery}, since these extrapolated points created a large scatter. Despite this, the general mass-recovery trend still holds, with halo growth clearly reducing the $\mathrm{H_2}$ recovery times. 

\begin{figure}
    \centering
    \includegraphics[width=\linewidth]{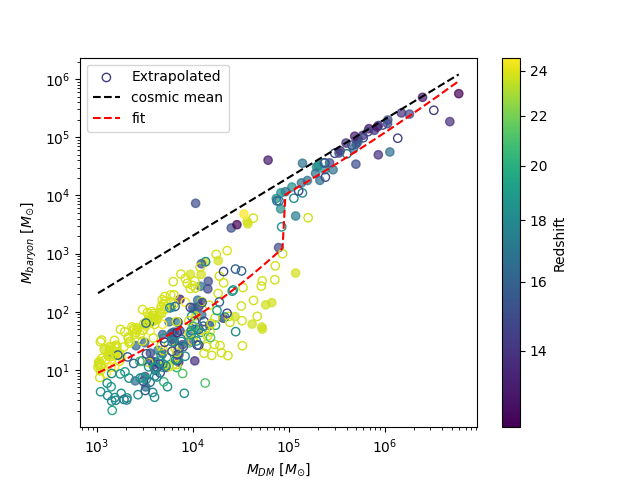}
    \caption[width=\linewidth]{Plot of baryon mass against dark matter for each of the halos that fulfilled the criteria. The cosmic mean is shown with a black dashed line and the fit to the data in red (equation \ref{eq:pw_eqn}).}
    \label{fig:dm_gas_ratio}
\end{figure}

\begin{figure}
    \centering
    \includegraphics[width=\linewidth]{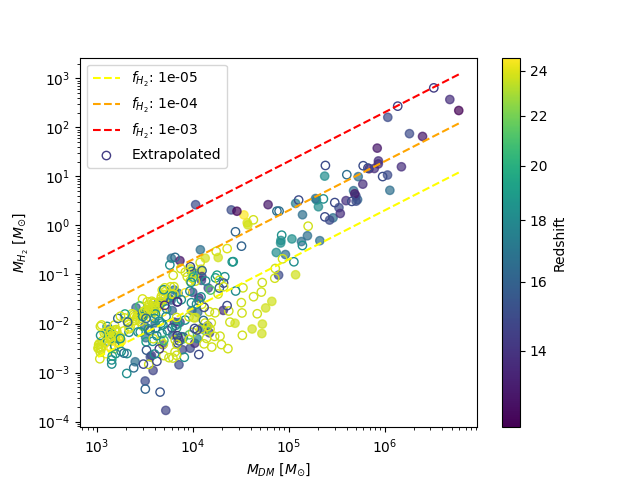}
    \caption[width=\linewidth]{Total molecular hydrogen mass inside the halos that met the criteria as a function of their dark matter content, measured just after a radiative event. Lines of different cosmic $\mathrm{H_2}$ fractions are shown with different coloured dashed lines.}
    \label{fig:mh2_Mdm}
\end{figure}

Figure \ref{fig:recovery} therefore presents a clear mass-recovery relation, with recovery times being reduced exponentially with increased halo mass. Halos with a dark matter mass of around $10^5$ $\mathrm{M_{\odot}}$ seem to experience an important turning point, as the scatter in recovery times is heavily reduced and limited to times below $\sim$30 Myr. As halos reach masses $ \ge 10^6$ $\mathrm{M_{\odot}}$, the overall trend falls to  $\sim10$ Myr. Since the recovery times reflect $\mathrm{H_2}$ production rates, we can understand that the first two stars to form did so in halos that had masses of $\sim$10$^5$ $\mathrm{M_{\odot}}$ because production rates at this threshold became efficient enough to reduce cooling times below Hubble time. From Figure \ref{fig:dm_gas_ratio} we also find that halos that reach this mass shift abruptly towards the cosmic mean baryon fraction, $\Omega_\mathrm{baryon} / \Omega_\mathrm{DM}$, being significantly baryon-poor at lower masses. By the time of the second star's radiation event, a larger number of halos in the system had reached or were approaching this threshold mass, leading subsequent star formation to occur within very small time delays. This is seen as the closely packed stars ($3^\mathrm{rd}$-- $11^\mathrm{th}$) in Figure \ref{fig:growth}. 

To further investigate the nature of the halos presented in the recovery graph, we looked at their baryon and molecular hydrogen masses. Figures \ref{fig:dm_gas_ratio} and \ref{fig:mh2_Mdm} show the baryon mass and $\mathrm{H_2}$ mass of the halos right before a radiation event. Figure \ref{fig:dm_gas_ratio} reflects an interesting behaviour in halo evolution, gas-poor halos dominate the lower masses and fall far below the cosmic mean, as would be expected in a $\Lambda$CDM universe. Once they approached the cosmological Jeans mass \citep[$\mathrm{M_{DM}}$ $\sim 10^5$ M$_{\odot}$,][]{2001PhR...349..125B}, they align with the cosmic mean and become potential Pop III star hosts. The important behavioural pattern highlighted in this figure is therefore the line break at $\mathrm{M_{DM}}$$ \sim 10^5$ $\mathrm{M_{\odot}}$, which coincides with the critical mass previously mentioned. We parameterised this behaviour by fitting a piecewise function to the data (seen in equation \ref{eq:pw_eqn}), where $x$ in this case is the dark matter mass of the halo, $f(x)$ the total baryon mass and $x_\mathrm{th}$ the threshold dark matter mass. 

\begin{equation}
    \label{eq:pw_eqn}
    f(x)= 
\begin{cases}
    Ax^{\alpha1},& \text{if } x \leq x_\mathrm{th}\\
    Bx^{\alpha2} + Ax^{\alpha1 - \alpha2},& \text{if } x > x_{th}\\
\end{cases}
\end{equation}

A total of 5 parameters were fit: $A=2.43$, $B=1.97$, $\alpha1=0.065$, $\alpha2=0.0824$ and $x_\mathrm{th} = 9.8\times10^4$ $\mathrm{M_{\odot}}$. From the fitted function we therefore find the break in power laws occurs at around $\mathrm{M_{DM}}$$\sim 1\times10^5$ $\mathrm{M_{\odot}}$ which is just below the cosmological Jeans mass: $\sim 1.5\times10^5$ $\mathrm{M_{\odot}}$. Using this, in conjunction with the previous recovery graph (Figure \ref{fig:recovery}) we see there are two distinct star formation epochs. At high redshifts, before a star has emitted any radiation in the system, gas poor minihalos dominate; and a race to meet the cosmic mean ensues. The first halos to approach the cosmic mean ($\Omega_\mathrm{baryon}\ /\ \Omega_\mathrm{DM}$) will be those that achieve masses of $\mathrm{M_{DM}} \sim 10^5$ $\mathrm{M_{\odot}}$ and are able to subsequently bring about collapse. Once the system has experienced a few radiation bursts, in our case two star formation events, star formation is suppressed. The halos that are aligned with the cosmic mean but have not yet formed a star experience a consistent photo-dissociation and $\mathrm{H_2}$ production cycle until their recovery times become smaller than the times between star formation. This growth and feedback relation is what forces the next set of stars to form within halos that grow to $\mathrm{M_{DM}} \sim 10^6$ $\mathrm{M_{\odot}}$, a whole order of magnitude above the initial $\mathrm{M_{crit}}$. Figures \ref{fig:dm_gas_ratio} and \ref{fig:mh2_Mdm} also explain the lower left clump of halos found in Figure \ref{fig:recovery}, these seem to be very gas and  $\mathrm{M_{H_2}}$ poor halos that were included due to our criteria only considering $\mathrm{{H_2}}$ fractions inside halo cores, and not the overall amount.

Continuing the analysis of these halos with Figure \ref{fig:mh2_Mdm}, we see that baryon mass is a good indicator of $\mathrm{H_2}$ content, as larger halos shift towards higher constant $f_\mathrm{H_2}$ cosmic mean lines. These are simply calculated as:  $f_\mathrm{H_2} \times (\Omega_\mathrm{baryon}\ /\ \Omega_\mathrm{DM})$. Halos that reside around or above the  $f_\mathrm{H_2} = 10^{-4}$ line are at the star formation stage, whereas halos below this would still need some time to produce more $\mathrm{H_2}$. The break around $\mathrm{M_{DM}} \sim 1\times10^5$ $\mathrm{M_{\odot}}$ is less evident in this figure, but still present since no halos above this mass fall below the $f_\mathrm{H_2} = 10^{-5}$ line. We also find that despite larger halos having a higher baryon mass content, they don't necessarily have a large enough amount of molecular hydrogen to form a star at that point in time. This is because these halos exist mainly at lower redshifts (z $<$ 17), and are experiencing successive star formation events that consistently lower their overall $\mathrm{M_{H_2}}$. These larger halos are therefore expected to quickly move up towards the $f_\mathrm{H_2} = 10^{-4} - 10^{-3}$ area during their recovery phase in order to meet the conditions for star formation. Given that these halos were observed right after star formation events, we are also able find the potential star-hosting candidates at each redshift using Figure \ref{fig:mh2_Mdm}.

By examining the recovery times, baryon mass and molecular hydrogen content together we are able to understand the  critical masses around which star formation becomes available to halos. We are also able to see at what mass scales and redshifts feedback delays are reduced to a minimum. Given that halos above $2\times10^6$ $\mathrm{M_{\odot}}$ move permanently above the $f_{H_2} = 10^{-4}$ cosmic mean and their recovery times (Figure \ref{fig:recovery}) remain well below 20 Myr, we can see that such halos in our system are able to recover $\mathrm{H_2}$ fast enough and contain a high enough content of $\mathrm{M_{H_2}}$ already to efficiently form a star after a radiation event. We can also infer that at $\mathrm{M_{vir}} \simeq 10^7$ $\mathrm{M_{\odot}}$, halos have short recoveries ($\sim 10$ Myr). It should be stressed however that these recoveries are calculated from the moment that a star dies in the system and no halo in our simulation forms a star while another is already present. In other words, we find the recovery process can't counteract LW feedback directly and even the larger halos in the simulation take time after a star has stopped emitting to produce enough $\mathrm{H_2}$ to cool efficiently.

\subsection{The Role of Major Mergers}\label{sec:mergers}
Having looked at radiative events, we explored other environmental factors that could affect star formation. We focused on mergers within a range of mass ratios in the main progenitor lines of our star forming halos. 

\begin{figure}
  \includegraphics[width=1\linewidth]{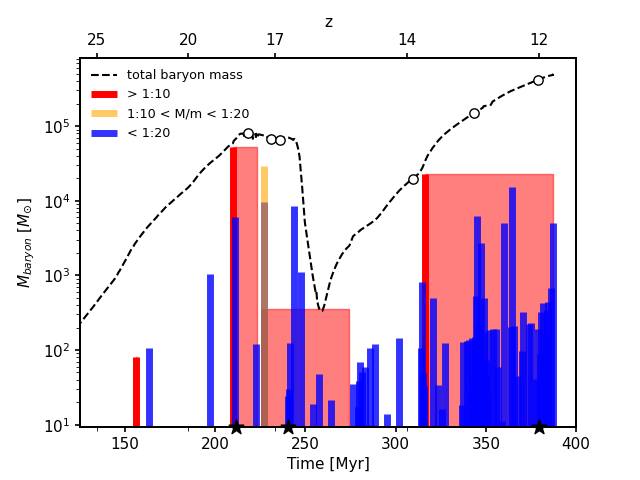}
  \caption{Baryon mass introduced from a range of merger types throughout Halo 3's evolution. The red bars are placed at the time the halos meet and are extended in a light red up to where the halo finder starts treating them as the same object. Note that the red bars are placed behind the yellow and blue bars and can become partially covered, such as is the case with the third major merger at $t=220$ Myr. The $2^\mathrm{nd}$, $3^\mathrm{rd}$ and  $12^\mathrm{th}$ star formation times are marked with black stars on the x-axis. The white dots on the dashed curve show the times at which the panels in Figures \ref{fig:particle_3} and \ref{fig:h12_merger} were taken.}
  \label{fig:mergers3}
\end{figure}

\begin{figure}
  \includegraphics[width=1\linewidth]{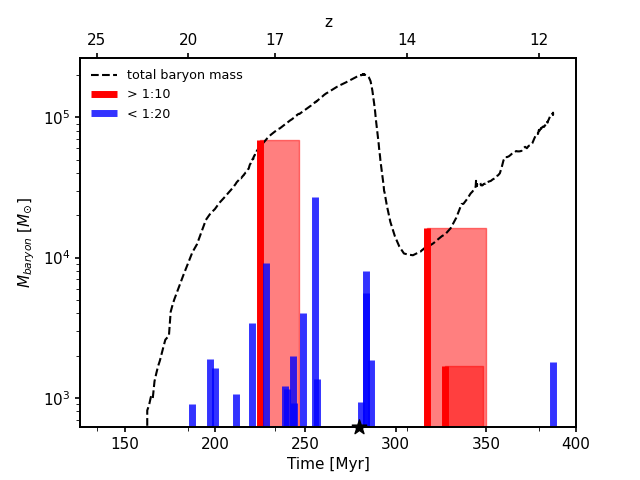}
  \caption{This is the same type of plot as Figure \ref{fig:mergers3} but for Halo 5/6. The time of the $5/6^{th}$ star's formation is marked with a black star on the x-axis.}
  \label{fig:mergers5}
\end{figure}

\begin{figure}
  \includegraphics[width=1\linewidth]{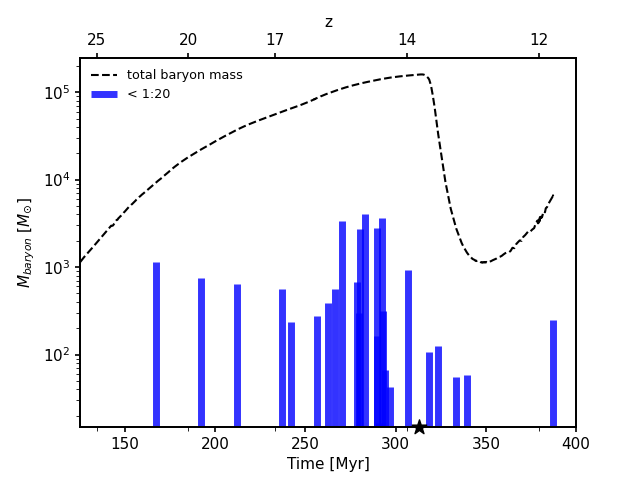}
  \caption{This is the same type of plot as Figures \ref{fig:mergers3} and \ref{fig:mergers5} but for Halo 9. The time of the $9^{th}$ star's formation is marked with a black star on the x-axis.}
  \label{fig:mergers9}
\end{figure}

Merger types are generally divided into two categories: major mergers, which are often restricted to ratios of 5-1:1 and minor mergers, that lie at a lower range of 10-5:1. However, due to the high resolution of the simulation, we extended these limits down to 20:1 to account for the prominence of small halos in our system. Even smaller mergers were also included in the analysis, with the smallest merging halos having masses of $900$ $\mathrm{M_{\odot}}$ since it is around and below this mass that the halo objects become poorly defined, making measurements increasingly unreliable. This categorisation of mergers resulted in a very small number of minor mergers being found. Potentially expanding the mass regime of these mergers down to 25-30:1 would lead to minor mergers being a more common event; however, the purpose of this particular analysis was to show the gas contribution of different merger types, not their statistical relevance. We found that the contribution of many small halos can lead to a considerable reintroduction of gas, although a single major merger is enough to outweigh their importance in the recovery process of halos. 

\begin{figure*}
  \includegraphics[width=1.0\textwidth]{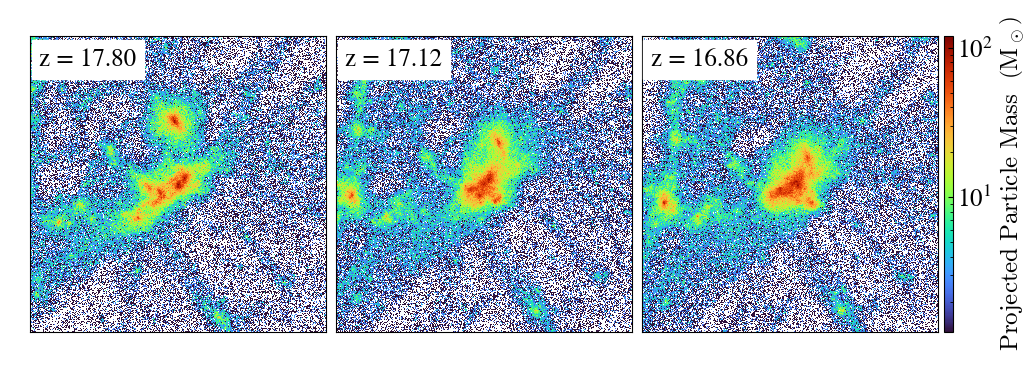}
  \includegraphics[width=1.0\textwidth]{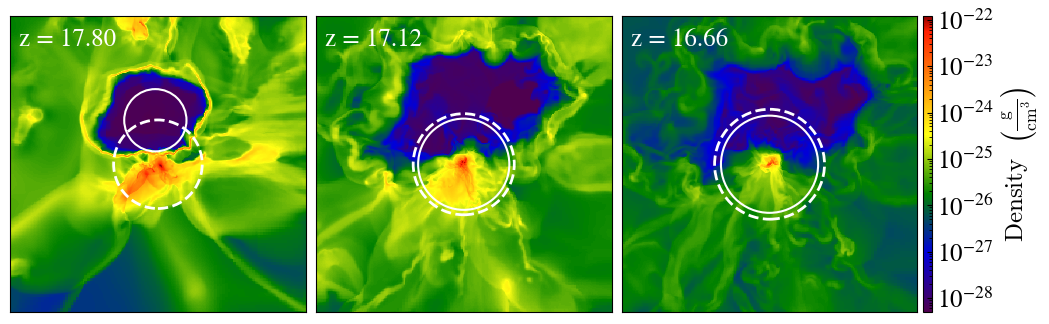}
  \caption{\textbf{Above}: Dark matter particle projections centered on Halo 3 along the x-axis. The panels show its merger with the recently evacuated Halo 2. The plot scale is of 1kpc and the depth of the projection is 1kpc. Halo 2 is observed to move in from the top of the plots.
  \textbf{Below}: Gas density slice plot centered on Halo 3 along the x-axis, the panels correspond to the particle projections shown above. The times chosen for the panels correspond to the time before the halo cores merged ($z=17.8$), the time during the merger ($z=17.12$) and the moment before star formation ($z=16.66$). The full and dashed circles represent Halos 2 and 3's virial radii respectively. The effect of the second star's supernova blastwave is observed as Halo 2 moves into Halo 3. Note the halo finder algorithm places Halo 2 inside Halo 3 before the cores have settled as can be observed when comparing particle and density slices.}
  \label{fig:particle_3}
\end{figure*}

\begin{figure*}
\centering
  \includegraphics[width=0.8\linewidth]{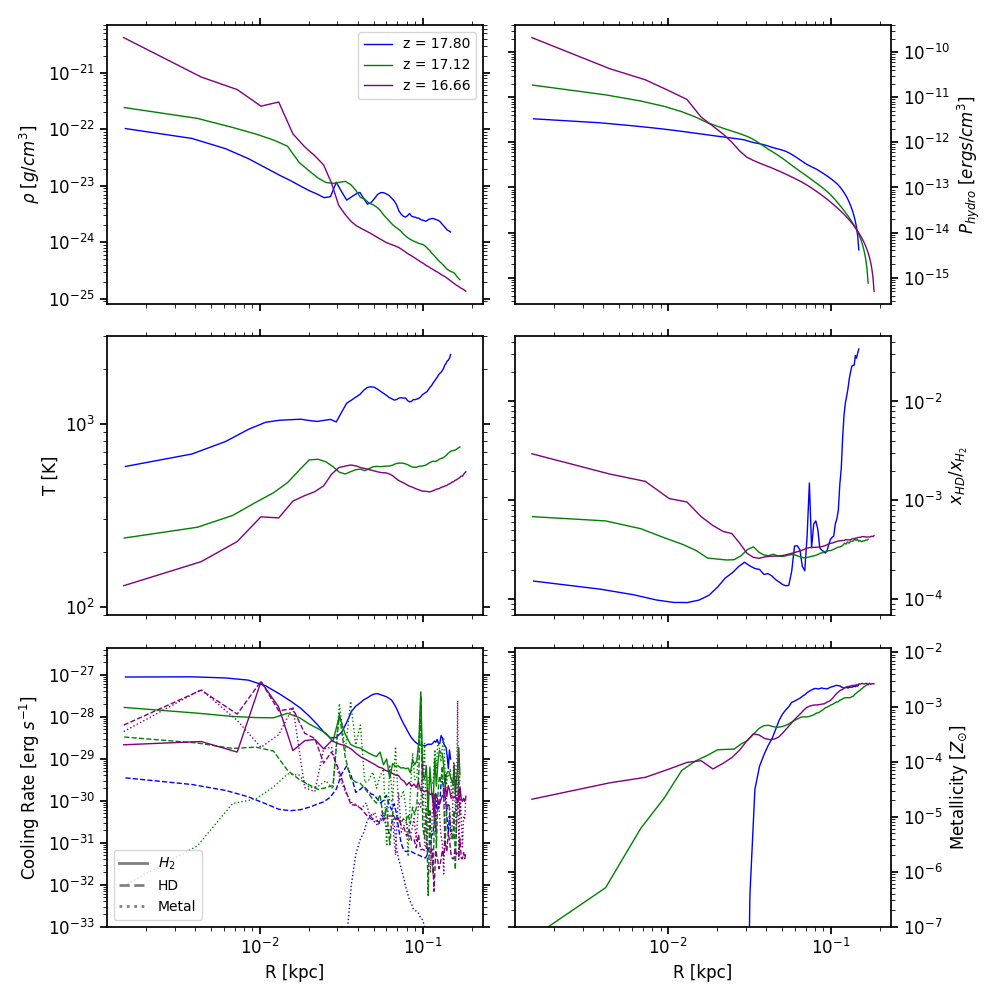}
  \caption{Mass-weighted, spherically averaged gas quantities: density, temperature, hydrostatic pressure, $x_\mathrm{HD}$/$x_{\mathrm{H_2}}$, cooling rate and metallicity; as a function of distance from the center of Halo 3. The cooling rate is further split into $\mathrm{H_2}$, HD and metals. The profiles are taken at different times around the merger event: pre-merger (blue), merger (green) and pre-star formation (purple). Note that because no HD photo-dissociation is implemented, the $x_\mathrm{HD}$/$x_{\mathrm{H_2}}$ ratio at times during and just after star formation is inaccurate, this effect is seen in the lower right graph, where a large spike at outer radii is observed in the highest redshift line (blue).}
  \label{fig:radial3}
\end{figure*}

In Figure \ref{fig:mergers3} we present the gas evolution as well as the merger history of Halo 3, past the formation of its first star up until the end of the simulation. We chose to analyse this halo more in depth specifically due to its unique evolution and formation history. Throughout the simulation it undergoes four major mergers, two of which occur around star formation times. The first star it forms (the third in the simulation) is seen as the dip in the dashed line at $\sim 255$ Myr in Figure \ref{fig:mergers3}. This is a direct effect of the stars' supernova depleting and photoevaporating most of the gas content inside the halo. After a period of around 100 Myr it then hosts the final star, soon after this, the simulation ends at $\sim$380 Myr. The coloured bars below the dashed curve show the total baryon mass that has directly entered Halo 3's progenitor line from merging objects at each timestep. For the smallest mergers recorded, in this case below the 20:1 regime but above the $900$ $\mathrm{M_{\odot}}$ mass limit, multiple objects tend to enter the halo within each snapshot, leading to the sometimes high cumulative gas contribution observed from the blue bars. It should also be noted that for the smallest objects, as well as early mergers, the bars mark the moment the merging object fell in and settled into the halo. Larger objects however, take longer to settle down, the ensuing process of dynamical mixing after the halos coalesce, leads to the formation of subhalo structures that first ``slosh" around before eventually falling into the core. This is represented as the extended bars in the plots, with larger halos taking longer to settle, as can be seen from the increasingly long shaded regions in Figures \ref{fig:mergers3} and \ref{fig:mergers5}. The halo finder considered phase-space occupation, not just physical-space locality, which allowed halos to persist beyond the point at which they are spatially merged. Hence, we considered the mergers to be complete once the halo finder took both objects to be the same. 

Examining further halo merger histories provides insight into role of major mergers and gas recovery. As seen from Figure \ref{fig:mergers3}, recently evacuated halos can quickly recover gas via major mergers. Another example of this is seen in Halo 5's evolution and merger history. From Figure \ref{fig:mergers5}, we find a similar gas recovery history as Halo 3, with Halo 5/6  also experiencing a set of major mergers that introduce large amounts of gas soon after its own star and subsequent supernova depleted it of its gas contents. This large accretion event meant it was ready for star formation by the end of the simulation. This greatly contrasts Halo 9, that undergoes no major mergers and has a final total baryon mass that is an order of magnitude lower than Halo 3 and 5/6 (Figure \ref{fig:mergers9}). Major mergers in these cases can therefore accelerate metal-enriched star formation by quickly reintroducing gas back into evacuated minihalos. Despite the rarity of the event (when compared to smaller mergers that occur consistently), a single major merger during a halo's evolution can re-introduce up to 100\% of the halo's gas content, as is seen in the merger histories for Halos 3 and 5/6. These events are also important in the recovery process after star formation, since we find they can introduce over 10\% of the halos previous peak gas content, accelerating recovery and later star formation.

\subsection{Mixed Mergers and Star Formation}

\begin{figure*}
\centering
  \includegraphics[width=1.0\textwidth]{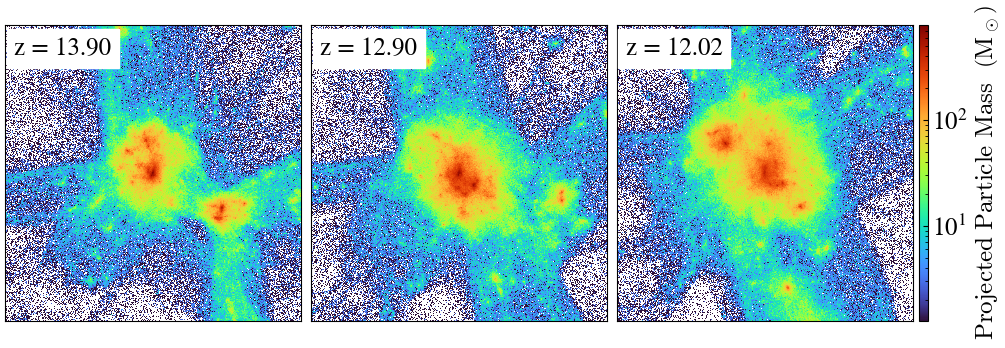}
  \includegraphics[width=1.0\textwidth]{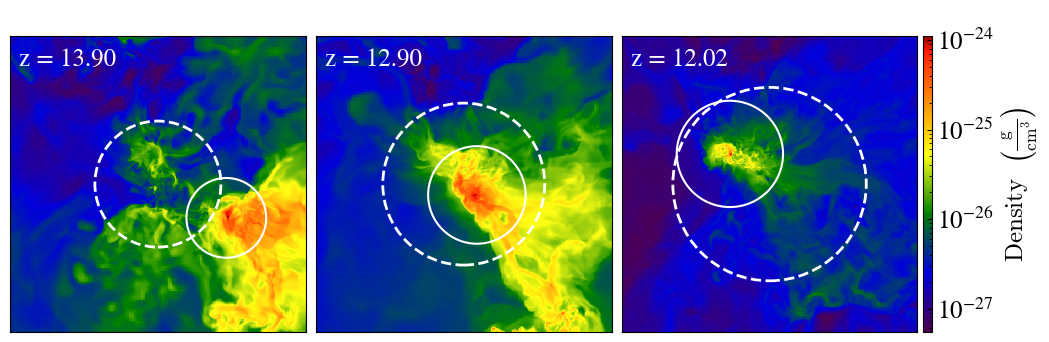}
  \caption{\textbf{Above}: Dark matter particle projection centered around Halo 2 showing the merger of halo halo full of gas with the recently evacuated Halo 3. The core of the incoming halo can be seen entering Halo 3 from the lower right corner. It passes through Halo 3's center and continues towards the upper left corner. The times chosen for the panels follow the same idea as those in Figure \ref{fig:particle_3} with a slightly larger plot scale of 2 kpc and depth of projection of 1 kpc.
  \textbf{Below}: Gas density slices corresponding to the particle projections shown above. The overplotted dashed and full line circles correspond to Halo 2 and the merging halos' virial radii respectively.}
  \label{fig:h12_merger}
\end{figure*}

\begin{figure*}
    \centering
    \includegraphics[width=0.8\linewidth]{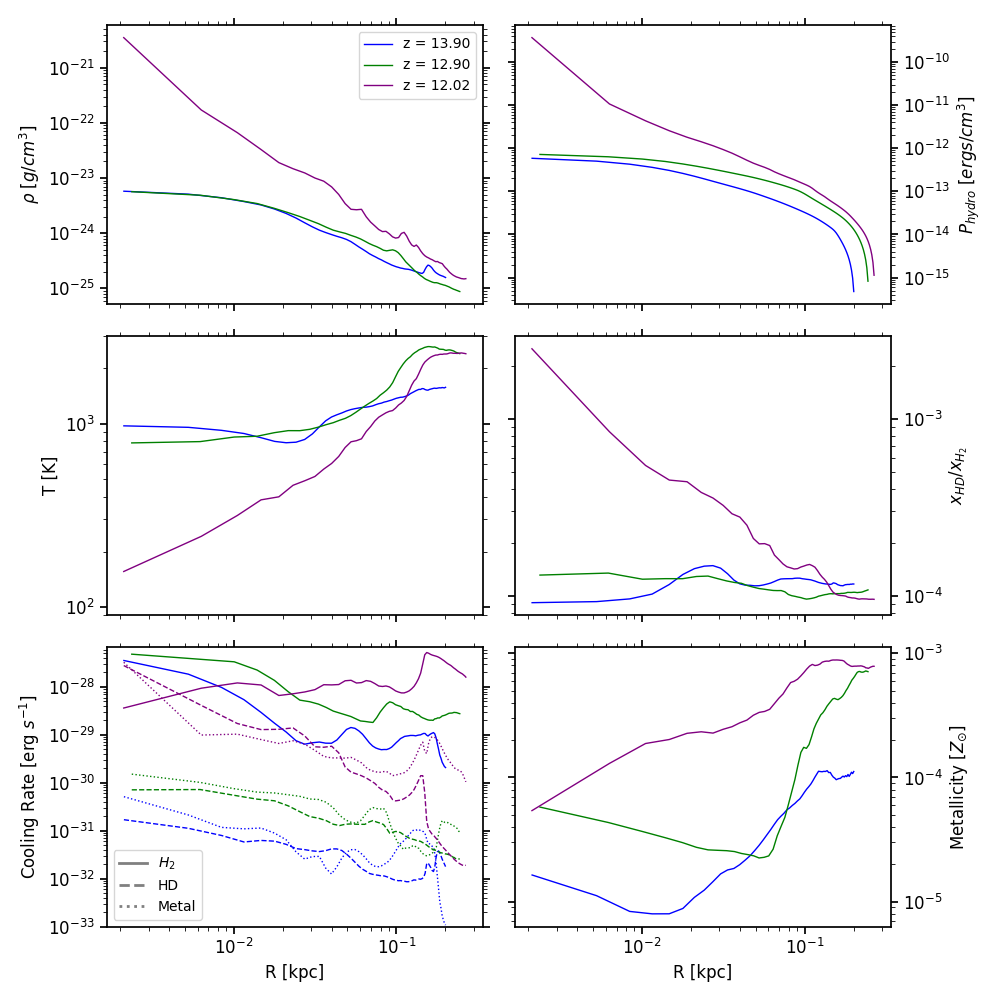}
    \caption[width=0.5\linewidth]{Mass-weighted, spherically averaged quantities: density, temperature, hydrostatic pressure and $x_\mathrm{HD}$/$x_{\mathrm{H_2}}$; as a function of distance from the densest point in the halo merging into Halo $12^*$. Times were taken pre-merger (blue), during merger (green) and pre-star formation (purple).}
    \label{fig:radial12}
\end{figure*}
 
During our analysis of Halo 3's merger history, we found the major merger seen starting at $\sim225$ Myr and ending at $\sim$275 Myr (in Figure \ref{fig:mergers3}) was an event that involved Halo 2. We can observe their proximity by the direct effect of Halo 2's supernova on Halo 3 as the flattening of the dashed curve. We better illustrate this merger with particle projection plots and gas density slice plots in Figure \ref{fig:particle_3}. Halo 2 first approaches Halo 3 at around $z=17$, just before these objects make contact, the second star forms and the shockwave from its supernova causes visible disruption inside Halo 3 (Bottom Figure \ref{fig:particle_3}). Halo 2 then ``sloshes" inside Halo 3 for around $50$ Myr, during which the third star is formed. This merger event is well within the classical regime that defines major mergers, with Halo 2 and 3 having a dark matter mass ratio of $\sim$1:4, yet we surprisingly find no delay in star formation. We attribute this to the evacuated nature of Halo 2 and define it as a ``mixed" merger in analogy with low redshift galaxy mergers.

The radial profiles in Figure \ref{fig:radial3} illustrate the physical changes observed in Halo 3's gas density, HD fraction, hydrostatic pressure, temperature, metallicity and cooling rates for different times throughout its merger with Halo 2. These profiles show an increasingly favourable set of conditions for collapse all while Halo 2 continues to move inside Halo 3. An increase in HD fraction is paralleled by a consistent drop in temperatures and increase in HD cooling, illustrating how this type of cooling become more present at the merger occurs. It should be noted that although HD photo-dissociation is not included in the simulation, the absence of nearby star formation occurring during the mixed merger and subsequent collapse means this does not affect our results. Another feature of the mixed merger that is not unexpected is the presence of metals in the low density gas surrounding the two merging halos. These have been expelled during the supernova associated with the now gas-poor halo. As Figure \ref{fig:radial3} (bottom, right) shows, some of these are able to mix into the gas-rich halo. However, the bottom left panel of Figure \ref{fig:radial3} also shows the cooling from metals to be negligible until just prior to star formation, and hence not a contributing factor when the gas cools the most. Thus, we maintain that the rise in HD in a notable feature of these mixed mergers.

Noting that electron fraction is high in relic HII regions, we understand that a higher HD fraction will be observed in halos that have already hosted a star. However, in this case, because Halo 3 is so close to Halo 2 during its SNe event, the increase in HD is also attributed to the large amount of turbulence caused by the blastwave). From Figure \ref{fig:radial3} we find that as the merger occurs, the cooling from HD quickly rises with cooling from metals only contributing near the end of the merger. Within its densest cell (where the star particle will eventually be placed) Halo 3 bottoms out at a temperature of around $78$ K, far below the limit of $\mathrm{H_2}$'s LTE. The continued plummeting of gas temperatures may seem surprising since the merger with Halo 2 is still ongoing, however, the evacuated state of Halo 2 minimizes the shock heating that would occur from gaseous collisions. This unique scenario means the presence of Halo 2 inside Halo 3 is ineffective at shock heating the gas and instead helps create a favourable environment for collapse. Hydrostatic pressure also increases as the dark matter mass content rises with the presence of Halo 2. More specifically, it ends up being about an order of magnitude higher than before Halo 2 entered Halo 3. Such mass growth also helps deepen the central potential well and create a denser core which maxes out at $10^{-21}$ $\mathrm{g\ cm^{-3}}$, a peak around which runaway collapse is triggered. 

After the formation of the third star, the newly evacuated Halo 3 (henceforth refered to as Halo $12^*$ since it forms the final star) remains gas-poor for close to $100$ Myr, until $z\sim$13.8, when a halo of $\mathrm{M_{DM}}\sim7\times10^5$ $\mathrm{M_{\odot}}$ and $\mathrm{M_{baryon}}\sim10^4$ $\mathrm{M_{\odot}}$ enters its virial radius. This event leads to the formation of the final Pop  III star, although soon after its birth the simulation ends. The star forms within the subhalo within Halo 3, as the merger does not finish before runaway collapse occurs. As seen from the dark matter and gas density projection plots in Figure \ref{fig:h12_merger}, the smaller halo moves into the virial radius of Halo 3 continuing past its core onto the other side. The dynamics of this interaction are quite interesting since the merging halo brings with it a significant amount of gas as the merging halo is at the star-formation stage. This particular merger is slow, with the subhalo taking roughly $40$ $\mathrm{Myr}$ to make its first crossing through Halo $12^*$. The overall timescale of the merger is uncertain since the simulation ends before the subhalo can settle down at the core. However, we can establish it takes over 80 Myr for the merging halo to settle. Such prolonged mixing is expected due to the massive nature of both these halos, roughly $\mathrm{M_{DM}} \sim 9\times10^5$ $\mathrm{M_{\odot}}$ and $3\times10^6$ $\mathrm{M_{\odot}}$ at the time of initial contact. When considering the free-fall and sound-crossing times ($t_{ff} = \sqrt{3\pi/32G\rho}$ and $t_{cs} = R\sqrt{m_h/\gamma k_b T}$ where $\gamma$ is the adiabatic index and $k_b$ the Boltzmann constant) of the larger merging halo at $z=13.9$, we find these are 25 Myr and 37 Myr respectively, further highlighting the length of this merger.

To further establish the parallels of this mixed merger with the previous one and assert the nature of these particular events, we again analysed the specific changes that occurred to the gas quantities before collapse. Following a similar analysis to that made with Halo 2 and 3's radial profiles, profiles for this event, shown in Figure \ref{fig:radial12} were produced. It is important to note these are centered around the merging halo, not Halo $12^*$. We again find HD enhancement occurring alongside a temperature drop and increased central pressure and densities. Although in this case there was a smaller enhancement of the HD fraction, something that is expected since Halo $12^*$ had no gas-rich major or minor mergers nor SNe blastwaves that could provide any prior turbulence. We therefore attribute the increase in HD to be caused by the long merger timescale. 
We find that metal cooling contributes at a similar level to HD early on as the metallicity is higher here than at analogous times for Halo 3. However, by the final time shown, HD cooling is higher than metal cooling throughout most of the inner halo despite a higher metallicity than in Halo 3. Hence, while present, the metals are not necessary to drive the cooling of the gas. The smaller HD fraction increase in this halo may also indicate that only a relatively small amount of HD production is required for HD cooling to take over from its less efficient precursor. The large mass scales involved in this merger are also well reflected in the hydrostatic pressure curves, as the core pressure increases by just under two orders of magnitude. Accordingly, the core densities rose to a central peak of $10^{-22}$ $\mathrm{g\ cm^{-3}}$ and would proceed to increase to $10^{-21}$ $\mathrm{g\ cm^{-3}}$ before the final collapse.

We therefore find that mixed mergers present a unique scenario in early star formation. The evacuated nature of one of the merging halos reduces the gaseous collisions that would normally stir turbulence and cause significant shock heating. The presence of a gas-poor halo, however, still becomes evident as central densities and hydrostatic pressure increases, accelerating $\mathrm{H_2}$ production and creating more optimal conditions for collapse. Such events are not a rarity in our simulation, since they occur multiple times in our small sample of halos. What is crucial with these types of mergers, however, is the timing. An example of a major merger that ended up becoming a dry merger due to its timing is seen with the Halo 4 and 5/6 pair. From Figure \ref{fig:mergers5}, we find that Halo 5/6 undergoes a merger event with another very gas-poor halo, this is Halo 4. (It can be seen in Figure \ref{fig:mergers5} as the small shaded red bar starting at $\sim$330 Myr with a height of $\sim 2\times10^3$ $\mathrm{M_{\odot}}$). The dry merger in this case occurs after Halo 4 and 5/6 have each formed their respective stars, leading this event to have no effect in accelerating collapse itself but instead results in fast gas accretion. This new, larger halo may have formed a star quite quickly had the simulation kept going. Despite this timing requirement, a system of early star-forming halos would be expected to undergo ``mixed" mergers frequently enough to take these events as potential avenues for collapse.

\section{Discussion}
\label{sec:discussion}
In this work we have analyzed the evolution of 9 minihalos that formed Pop  III stars in the \texttt{Pop2Prime} simulation. We focused on the environmental factors, in the form of radiative feedback and major mergers, that directly affected these halos' ability to effectively cool and form stars. As mentioned in Section \ref{sec:code}, the simulation analyzed here was the same as the high-resolution simulation in \cite{Smith2022}. This simulation was an extension of an earlier simulation presented in \cite{Britton2015} that set a lower threshold for the metallicity of Pop  III stars. They therefore found only 2 Pop III star-forming halos before metal-enriched collapse occurred and the simulation ended. These are Halos 1 and 2 in this work. Our higher threshold allowed the simulation to carry on longer, thus providing a larger sample of minihalos and enabling us to follow the evolution of halos long after they had hosted a star.

\subsection{Comparison with Previous Works}
We initially followed much of the established work on radiative feedback (\citealt{Machacek_2001}; \citealt{Abel2002}; \citealt{Yoshida2003}, \citealt{Reed_2005}; \citealt{Oshea2008}; \citealt{Greif2010}), and focused on the effect of LW radiation in isolation as a factor in delaying collapse. Because our simulation did not implement a uniform background field, instead turning point-source radiation on or off depending on the presence of a star, we were able study feedback as a discrete event. Despite this unique scenario, our findings are in close agreement with works that implement a constant background radiation; in the absence of LW photons, as is the case for Halo 1, we observe star formation occurring in halos with $\mathrm{M_{DM}} = 10^5$ $\mathrm{M_{\odot}}$, while repeatedly radiated halos undergo delayed collapse and do not form stars until reaching masses close to or above $\mathrm{M_{DM}} = 10^6$ $\mathrm{M_{\odot}}$.

We find quite strong agreement with the mentioned past works, though more recent studies 
such as \cite{Skinner_2020}, \cite{Schauer2021} and \cite{Kulkarni_2021} find slightly larger critical masses: $\mathrm{M_{crit}} \sim 2-3\times10^5$$\mathrm{M_{\odot}}$ for no radiation and $\mathrm{M_{crit}} \sim 1-3\times10^6$$\mathrm{M_{\odot}}$ with radiation. While \cite{Park_2021}, that implement an X-ray background, found that this effect could significantly lower a halos critical mass in weak LW backgrounds. We find however their results are still within reasonable agreement with our findings. Despite the different simulation setups between works; with other studies such as \cite{Schauer2021} and \cite{Kulkarni_2021} also including streaming velocities and self-shielding effects, as well as modelling radiation as a smooth varying background, the overall consensus on halo critical masses remains consistently around $\mathrm{M_{crit}} \sim 10^5$ $\mathrm{M_{\odot}}$

Unlike our work, many of the studies considering radiative feedback have also attempted to parametrize the evolution of the $\mathrm{M_{crit}}$. We have not done this here since our sample is so small. We instead compared predictions for $\mathrm{M_{crit}}$ from \citet{Kulkarni_2021} with our own observations and find quite reasonable agreement for multiple halo masses within a range of $\mathrm{J_{21}}$ intensities. As previously mentioned, under no radiative influence, \citet{Kulkarni_2021} predict a halo mass that is in good agreement with Halos 1 and 2 here. Halo 4's star formation mass is in agreement with their prediction for $\mathrm{M_{crit}}$ when $\mathrm{J_{21}} = 1$ and Halo 9 and 10/11 sit right above and below this fit. Halo 9 experienced no major mergers before star formation and so was subject only to radiative delays. Having an $\mathrm{M_{crit}}$ that falls near the predicted masses for $\mathrm{J_{21}} = 1$ may imply that under radiation effects alone, halos will form stars at masses that follow this curve. Halo 4 and 10/11 both experienced a major merger event but with halos that were far below cosmological Jeans mass, with the largest halos they merged with having $\mathrm{M_{DM}} = 8\times10^3$ $\mathrm{M_{\odot}}$ and $3\times10^4$ $\mathrm{M_{\odot}}$ respectively. In contrast Halos 5/6 and 7/8 experienced major mergers with much larger halos that were close to the cosmological Jeans mass, making these events more disruptive since the halos involved had enough gas to potentially form a star. We therefore find that the cumulative effect of radiation and major mergers can steepen a halos $\mathrm{M_{crit}}$ curve. Halo 3 is an extreme example of this since it undergoes major mergers, hosts SNe blastwaves, experiences radiative feedback, and forms the final star in the simulation at an $\mathrm{M_{crit}}$ that is nearly two orders of magnitude above the first halo, aligning with the $\mathrm{J_{21}} = 10$ curve. This further implies that dynamical feedback can produce $\mathrm{M_{crit}}$ curves that align with more extreme background intensities. In order to accurately parameterize $\mathrm{M_{crit}}$, the contribution from mergers and SNe blastwaves should be considered.

We also centered our analysis on $\mathrm{H_2}$ recovery and its relation to halo growth, as this directly dictates whether efficient cooling and therefore collapse will take place. We find, as is expected from the analytic models in \citet{Tegmark1997}, that radiative feedback becomes less effective as halos grow. With recovery times decreasing exponentially with increased halo mass and halos above $10^6$ $\mathrm{M_{\odot}}$ tending to have recovery times that remain below 30 Myr. We also predict that larger halos reach a minimum recovery time around 10 Myr, with larger halos potentially pushing this to even smaller times. We also find a crucial growth period, as halos approach $10^6$ $\mathrm{M_{\odot}}$, where star formation becomes increasingly accessible to a larger proportion of halos and collapse can only be brought about when recovery times become smaller than the competing halos' collapse times. As we do not observe any instances of co-existing Pop III stars (apart from the double systems), we see how the LW radiation forces halos to repeatedly compete with each other in order to bring about collapse.

$\mathrm{H_2}$ recovery times are also heavily dependent on the gas content and total $\mathrm{H_2}$ mass within a halo. Understanding the global trends of these quantities during halo growth and comparing them to the cosmic mean provides a clearer picture on why certain halos were able to host the correct environment for star formation to be possible. In our simulation we find two clear epochs: at high redshifts, halos remain gas-poor until they reach masses close to the cosmological Jeans mass $\sim 10^5$ $\mathrm{M_{\odot}}$. At $z\ge17$ a sufficient number of halos have reached this critical mass for continual and successive star formation to take place, at this stage halos require cooling times that fall below the average collapse timescales. This leads to the $\mathrm{H_2}$ dissociation and formation cycle presented extensively in \citet{Tegmark1997} and is what pushed $\mathrm{M_{crit}}$ to $\sim$10$^6$ $\mathrm{M_{\odot}}$. These findings are in close agreement with all the aforementioned studies, although other factors such as self-shielding and baryon-dark matter streaming velocity have been shown to push this mass to even higher thresholds, $\sim 10^7$ $\mathrm{M_{\odot}}$ (\citealt{Trenti_2009}; \citealt{Schauer2021}; \citealt{Kulkarni_2021}).

\subsection{A New Major Merger Scenario}

We find and establish a novel merger type for Pop  III star forming halos in the form of ``mixed" mergers. These can trigger collapse in halos that would otherwise have taken longer to do so. Mergers between gas-poor galaxies, defined as ``dry'' mergers, have already been studied in early galaxy formation scenarios, since they enable significant mass growth without altering stellar populations (\citealt{Bell_2006}; \citealt{Khochfar_2006}; \citealt{Khochfar_2009}). We find a similar situation here, since halos that end up being depleted of their gas content after a SNe type II event maintain their dark matter content so have high central densities. They can therefore provide their gas-rich neighbours with a deeper potential well, causing the required increase in central densities to bring about collapse without the major disruptions that would be expected from regular mergers. \cite{Shchekinov_2005} suggested that under the correct conditions, halo mergers could boost the HD content in a halo because of the increased electron abundance in post-shocked gas. This enables a new and more efficient cooling mechanism to take over and lower temperatures to be reached. \citet{Norman_2010} addressed this cooling channel in Pop III star formation, establishing that relic HII regions, such as our recently evacuated halos, can form a higher amount of HD due to the gas in these halos forming from initial conditions that have been thermally and chemically altered. \cite{Prieto_2012} explored this idea, focusing on the effect of turbulence created by mergers increasing the HD content. \cite{Bovino2014} would also present similar results by following collapse to higher densities and showed significant ionization and HD enhancement from a major merger event. They found temperatures dropped down to $60-70$ K once HD cooling was triggered much like what we observed in the halos that undergo a mixed merger event. In contrast, our non-merging halo example, Halo 9, cools down to only $\sim$240 K.

These works agreed that shocks and turbulence from major mergers significantly boost $\mathrm{H_2}$ and HD in halos, enabling cooling to temperatures close to the CMB limit. However the sustained high turbulence measured from such mergers, nearly $10$ $\mathrm{km\ s^{-1}}$ in the case of \cite{Bovino2014}, alongside the significant shock-heating inside the gas cores, would delay star formation in halos where $\mathrm{H_2}$ cooling was still inefficient, as is the case for our minihalos. Although these past works aren't able to follow the evolution of the halos down to star formation and past this, they don't expect star birth to be triggered until after the mergers have finished. We instead find collapse occurs whilst the mixed mergers are still ongoing, making such events a direct catalyst for star formation. Mixed mergers are also not a rare occurrence in our simulation, though the timing of these is crucial for them to help bring about collapse. An example of this is the Halo 4 and 5 system, they exist close to each other ($\sim$0.3 kpc between their centers), however Halo 4 starts merging into Halo 5 after they have both hosted their corresponding stars. Making this dry merger redundant in the formation of Pop III stars but potentially very useful in the creation of subsequent stellar generations.

Despite our small sample size, we can still assess the potential impact of mixed mergers within the broader context of early universe star formation. In order to do this however, we need to also consider the nature of the simulation or model being analysed. Mixed mergers, like any other phenomenon that occurs due to the physics framework applied, cannot be isolated from the rest of the simulation. However, it is possible that low resolution simulations that do not adequately resolve the central potential in minihalos might not see star formation induced by such events. In terms of star formation rates (SFR), our small sample size again limits the conclusions we can make about the impact of mixed mergers. It does, however, leave open avenues for future work to build on. Mixed mergers may alter SFRs in denser regions compared to mean or void areas. Denser populated zones in the universe are expected to have a larger abundance of evacuated halos these could accelerate star formation early on. However, these regions also have smaller clustering lengths, leading halos to experience high levels of LW radiation from the first star forming halo. Whether mixed mergers have outweigh this effect would be an interesting avenue to take. Semi-analytic models based on merger-trees from N-body simulations \citep{magg_2022} could incorporate the phenomenon of mixed mergers to estimate its overall impact on SFRs.

Though mixed mergers are present in early star formation, they may have a limited timeframe within which they help produce Pop III stars. Because they are relic HII regions that have seen star formation, they undergo small amounts of enrichment. In our case, the close SNe encounter leads to additional metals remaining in Halo 3 and therefore Halo $12^*$. The nature of mixed mergers means halos involved in these experience small levels of metal enrichment and metal cooling. While metals are likely to be present in the environment of most mixed mergers, we have demonstrated that they are not explicitly necessary to enhance cooling and induce star formation, as this can be accomplished by HD and $\mathrm{H_2}$. What is less clear is how the metals will mix into the dense gas where the stars next will form. In both cases we study, the metallicity only marginally exceeds that necessary for fragmentation due to dust \citep{Schneider2012} and hence the possibility remains that star formation in the manner of Pop III (i.e., with no appreciable metal influence) occurs in these events. 

\subsection{Major mergers and recovery}
Another important result that should be highlighted, despite already being mentioned previously, is the ability major mergers have to inject large amounts of gas into halos. The notion of recovery via fallback of supernova ejecta presented in works such as \cite{Jeon_2014} is completely absent in over half of our halos, since they recover most of their gas via major mergers instead of smooth accretion after star formation. Defining the recovery time-scale as the time delay between a Pop III SN explosion and the replenishing of cold, dense gas within the halo, \cite{Jeon_2014} found that halos around $5\times 10^5$ $\mathrm{M_{\odot}}$ have recovery times of $90$ Myr. We find this is true for halos that recover gas via a major merger after star formation, however those that do not (such as Halo 9) barely reach half of their original maximum after this time. Major mergers are therefore a crucial component of gas recovery for halos in the $5\times 10^5 - 10^6$ $\mathrm{M_{\odot}}$ mass range. Minor mergers in this case play a less significant role in mass growth compared to major mergers. Aside from bringing in less baryon mass, their dynamical friction time scales are longer, making minor mergers impact halo mass growth at later times. We therefore find major mergers are a more dominant factor in the formation of structure in the early Universe.

\section{Summary \& Conclusion}
\label{sec:conclusion}
We analyzed a cosmological radiation hydrodynamics simulation run with the \ENZO AMR + N-body code to study the effects of LW radiation and major mergers on Pop III star formation. The reduced halo sample size along with the simulations' high resolution capacity allowed a detailed analysis of each star-forming halo and the interaction between these. Moreover, by not implementing a uniform background field and instead turning point-source radiation on and off during each star's lifetimes, we were also able to focus on the effects individual stars had in the system. We therefore focused heavily on the interplay of halo evolution and star formation to produce a comprehensive narrative of the creation of the first stars. Our final conclusion are summarized as follows:

\begin{itemize}
\item We found a clear relation between $\mathrm{H_2}$ recovery times and halo mass. With larger halos recovering faster and therefore experiencing increasingly smaller delays. Without radiation, the halo mass for star formation was, $\mathrm{M_{crit}} \sim 10^5\ \mathrm{M_{\odot}}$, this increased to roughly $10^6\ \mathrm{M_{\odot}}$ after only two radiative events. We expect to find that at a high enough halo mass, $\sim 10^7\ \mathrm{M_{\odot}}$, the delay caused by feedback effects reaches a minimum that is around 10 Myr.

\item We also established a scaling relation in the form of a piecewise function for the baryon mass and dark matter content inside minihalos, finding a break in the power law at $\mathrm{M_{DM}} \sim 10^5\ \mathrm{M_{\odot}}$. This inflection point coincides with the cosmological Jeans mass, which is the mass around which previously gas-poor halos move towards the cosmic mean.

\item We presented a novel major merger scenario that can accelerate Pop III star formation and define these as ``mixed" merger events. These catalysts for star formation can provide the correct local environmental changes to bring about collapse while avoiding the detrimental effects associated with major mergers at low redshift. The statistical relevance of such mergers as well as a closer look at the halo dynamics at play would be an interesting direction to take in future studies.
 
\item We additionally found major mergers as important processes for gas recovery in halos, being especially important in the recovery of halos that have previously hosted a star and are in a depleted state. We find that major mergers can introduce up to 100\% of the halo's current gas content before star formation and can reintroduce up to 10\% of an evacuated halos previous maximum gas mass.
\end{itemize}

In this work we have attempted to clarify the series of events that lead to Pop III star formation. Focusing on the radiative effect of individual stars, we reached similar conclusions to past works that implemented a background radiation; while also allowing us to focus on a different feature resulting from feedback effects, the recovery of a halo's molecular content. Our findings fit well within the established theories and we have attempted to further deepen the understanding of the radiation and recovery process by providing insight into the relations between the baryon and dark matter content and $\mathrm{H_2}$ mass of these early minihalos. The high resolution of the simulation also allowed us to get a closer look at halo-on-halo interaction, with mixed mergers being our novel addition to the major merger narrative. While major mergers may sometimes be effective suppressants of star formation, we have shown that there is more to their role in early star formation than has yet been presented in works so far.

\section*{Acknowledgements}
For the purpose of open access, the author has applied a Creative Commons Attribution (CC BY) licence to any Author Accepted Manuscript version arising from this submission. We thank the anonymous referee for their comments, which greatly improved the manuscript. BDS and SK are supported by STFC Consolidated Grant RA5496. JHW is supported by NSF grants OAC-1835213 and AST-2108020 and NASA grants 80NSSC20K0520 and 80NSSC21K1053. BWO acknowledges support from NSF grants 1908109 and 2106575 and NASA grants NNX15AP39G, 80NSSC18K1105, and 80NSSC21K1053. The \texttt{Pop2Prime} simulations  were performed on Blue Waters, operated by the National Center for Supercomputing Applications with PRAC allocation support by the NSF (award number ACI-0832662).

\section*{Data Availability}
The data underlying this article will be shared on reasonable request to the corresponding author.



\bibliographystyle{mnras}





\bsp	
\label{lastpage}
\end{document}